\documentclass[usenatbib]{mn2e}
\usepackage{txfonts}
\usepackage[dvips]{graphics}
\title[Triggered star formation in Bright Rimmed Clouds: The Eagle Nebula revisited]{Triggered star formation 
in Bright Rimmed Clouds: The Eagle Nebula revisited}
\author[J. Miao, Glenn J. White, R. Nelson,M. Thompson, L.Morgan]{J.Miao$^{1}$\thanks{E-mail: 
j.miao@kent.ac.uk}, Glenn J.White$^{1}$, R.Nelson$^{2}$, M.Thompson$^{3}$, L.Morgan$^{1}$   \\
$^1$Centre for Astrophysics \& Planetary Science, School of Physical Science, University of 
Kent, Canterbury, Kent CT2 7NR, UK \\
$^2$School of Mathematical Sciences, Queen Mary, University of London, Mile End Road, London E1 4NS, UK\\
$^3$School of Physics Astronomy \& Maths, University of Hertfordshire, College Lane, Hatfield, AL10 9AB, UK}
\begin{document}

\date{Received date / Accepted date}

\pagerange{\pageref{firstpage}--\pageref{lastpage}} \pubyear{2005}

\maketitle
\label{firstpage}

\begin{abstract}
A  three dimensional Smoothed Particle Hydrodynamical (SPH)
model has been extended to study 
the radiative driven implosion effect of massive stars on the dynamical evolutions of surrounding 
molecular clouds. The new elements in the upgraded code are the inclusion of 
Lyman continuum  in the incident radiation flux and the treatment of  hydrogen ionisation process; 
introducing ionisation heating \& recombination cooling effects; and adding a proper description of 
the magnetic and turbulent pressures to the internal pressure of the molecular cloud. 
This extended code provides a realistic model 
to trace not only  the dynamical evolution of 
a molecular cloud, but also can be used to model 
the kinematics of the 
ionisation \& shock fronts and the photo-evaporating gas surrounding
the molecular cloud, which the previous code is unable to deal with.  

The application of this newly developed model to the structure of 
the middle Eagle Nebula finger suggests that 
the shock induced by the ionising radiation at 
the front side of the head  precedes an ionisation front moving towards 
the center of the core, and that the core at the fingertip is at transition stage evolving toward 
a state of induced star formation.   
The dynamical evolution of the velocity field of the simulated cloud structure is discussed
to illustrate the role of the self-gravity and the different cloud morphologies 
which appear at different stages 
in the evolutionary process of the cloud. 
The motion of the ionisation front and  
the evaporating  gas are also investigated. The modelled gas evaporation rate
is consistent with that of current other models and the density, temperature and chemical profiles 
are agreement with the observed values. 

The relative lifetimes of different simulated cloud morphologies suggests a possible 
answer to  the question of why
 more bright-rimmed clouds are observed to possess a flat-core than an elongated-core morphology.
\end{abstract}

\begin{keywords}
star: formation -- ISM: evolution -- ISM: HII regions -- ISM: kinematics and dynamics -- radiative transfer.
\end{keywords}

\section{Introduction}

The Eagle Nebula (M16) presents an excellent laboratory in which to study 
the influence on star formation due to the 
presence of nearby massive stars, which can lead to radiative implosion of 
surrounding molecular cores and the initiation of secondary star formation \citep{Bertoldi, Lefloch}.
The intriguing  structures in 
the heads of the Elephant Trunks in the Eagle Nebula have inspired many studies.
A comprehensive survey conducted by  \citet{Hester} using  the HST 
Wide Field and Planetary Camera (WFPC2) resolved a number of  Evaporating Gaseous Globules (EGGs)
at the tip of finger-like features protruding from  columns of cold gas and dust 
of the Eagle Nebule \citep{Hester}. More investigations  on    
the physical characteristics and the future evolution of the head structures have also been carried
out since then \citep{White,Williams,Fukuda,Sugitani,Thompson,Urquhart}. 

In one of our early papers \citep{White}, we reported  molecular line, 
millimetre/submillimetre continuum, 
and mid-IR observations of this region. The steady state of the head structure of 
the middle finger of the Eagle Nebula 
was estimated  using a SPH model which included a simple treatment of
an isotropic FUV radiation field  
($6.5 < h\nu < $ 13.6 eV).  The comparision of the estimated steady state of the 
head structure of the middle 
finger  with the observed data
 suggested that the FUV ionisation induced shock 
might have not passed the core at the middle fingertip, so that 
the head structure of the fingertip could not be taken as the 
remains of radiatively driven implosion.  
The observations revealed that the core structure shows the 
characteristics similar to those expected  in the earliest stages of protostellar 
formation, therefore it was suspected that the head structure of the middle finger in
the Eagle Nebula might evolve from  
 a large and dense structure which existed prior to the expansion of the HII region. 

Recently there have been other observations that have supported the idea of triggered protostellar
formation at the tip of the middle finger, following the interaction 
of the natal  molecular cloud with the ionising radiation from the nearby OB stars 
\citep{Fukuda,Mcgaughrean,Sugitani,Thompson}. 
The question is whether or not  UV radiation from 
the nearby massive stars could trigger the collapse of the head structures in Eagle Nebula
and lead to the next generation of star formation.      
The challenge is to develop theoretical models that provide a reliable description of the dynamical 
evolution of these clumpy structures and which can trace back the history  of the clumps,    
 accommodate the current observations, and predict the future evolution of 
these fingertip structures.   

A precise description of the genesis 
and the future evolution of the head structures such as those of the Eagle Nebula fingers requires  
a more rigorous dynamical modelling which should include Lyman continuum in the external 
radiation field, since
hydrogen ionisation ought to play a dominant role in the dynamical evolution of the 
molecular cloud structures like those in  Eagle Nebula.    

A substantial amount of theoretical work has already been carried out to investigate 
the evolution of dense, gaseous clumps which are in the vicinity of massive OB
stars. These include the Radiatively Driven Implosion (RDI) mechanism \citep{Bertoldi} 
which was developed using an approximate analytical solution for the evolution of a 
spherical symmetric neutral cloud  
subject to the radiation of a newly formed nearby star; 
and the cometary globule model 
\citep{Lefloch,Leflocha}, which describes the 
dynamical evolution of 
a neutral globule illuminated by the ionising radiation of OB stars based on  
2-D hydrodynamical simulations. 
Although these 2-D models successfully reproduced some of the observed characteristics of 
 objects influenced by 
nearby massive stars, they    
did not adequately include self-gravity \citep{Bertoldi}, 
nor  self-consistently treat the 
chemistry and thermal evolution \citep{Lefloch,Leflocha}. Consequently 
 they are unable to deal with the details of triggered star formation by UV radiation. 

\citet{Williams} presented a 2-D (i.e., cylindrical 
symmetry of the cloud structure is used) hydrodynamical 
simulation of the  evolution of photoionised clumps with similar characteristics  to 
those found in the Eagle Nebula. The simulated results over a variety of initial conditions 
proposed a different evolutionary scenario from that of  \citet{White} 
for the head structures, suggesting that the head structures at the tips of  
the Eagle Nebula trunks may have already survived
the propogating shock induced by the external UV radiation field and are in a 
near-equilibrium state  about $10^5$ years later. In this modelling, self-gravity of the molecular
gas was also not included. 

The existing discrepancy in the descriptions of the structure of the heads at the finger tips    
and the recently reported new observations on the finger tips' structure suggests the necessity 
of revisiting  the issue of UV radiation triggered star formation at these fingertips. In order to do this, 
a self-consistent and comprehensive 3-D 
BRCs' model is required which could be used to study the
UV radiation triggered star formation in BRCs and which more generally can be used 
to describe observations in all of  BRCs, proplyds and HII regions, to meet  
with the ever-increasing interest in the investigation of the nature of BRCs
and proplyds in HII regions. 
It is obviously seen from the above that this fully 3-D BRCs model should be able to
 self-consistently and simultaneously 
treat the self-gravity, thermal evolution, radiative transfer and  chemistry. Therefore it is 
our intentin to develop the first self-consistent and comprehensive 3-D BRCs's model based
on a previous 3-D SPH code, which was originally develped to investigate the evolution of isolated 
molecular clouds under the effect of the interstellar medium radiation \citep{Nelson}.  
 
In the following sections in the present paper we will firstly present 
an outline for the previous 3-D SPH model and 
then introduce new elements neccessary for simulating the evolution of BRCs. 
The simulation results will be discussed and compared with observations, 
and  the evolutionary of the simulated cloud structure will be investigated. 
A comparision of the resuluts with those based on the previous code is made and then we infer the 
future evolution of the Eagle Nebula 
finger from its current state. Based on the good agreement between simulations and 
observations, a reasonable explanation for the observed unbalanced numbers 
between the flat-core and elongated-core BRCs is proposed.   
                     
\section{The three dimensional SPH modelling}
The physical properties of the head structure of the Eagle Nebula presented in our previous 
paper \citep{White} was based on the Smoothed Particle Hydrodynamics (SPH) code 
developed by \citet{Nelson}, which self-consistently  treats the  
self-gravitation, chemical and thermal evolution in the molecular cloud. 
Starting with a cylindrical finger of radius 0.1 pc with a spherical  molecular
cloud of 31 M$_{\sun}$ at its head,  which is 
under the effect of a FUV radiation field 
(with photon energy lower than 13.6 eV and enhanced from the +z-direction). Based on 
 an assumed static density distribution (i.e, under equilibrium assumption), the 
temperature and chemical distributions, after the FUV radiation is swiched on for $10^5$ 
years, showed similar characteristics  
to observations of the middle finger. However the detailed and accurate description on 
the whole dynamical evolution, especially on the genesis and future
evolution of the observed structures was not discussed at that time, since the original 
SPH code did not include the hydrogen ionising influence of the Lyman continum radiation ($ h\nu \ge 13.6$ eV),
the ionisation heating, recombination cooling or the relevant radiative tranfer process. These are clearly
 important since the effect of  the UV radiation of nearby massive stars
on their surrounding molecular clouds tells us that  hydrogen ionisation is one of the most important 
 physical processes in radiation driven implosion models  
\citep{Bertoldi,Lefloch,Williams}. 

Recently, \citep{Kessela, Kesselb}
have developed a three-dimensional SPH model of the
radiation-driven implosion of molecular cloud which includes both self-gravitation and hydrogen ionisation 
from nearby stars. Their model provided the first 3D SPH code to treat ionising radiation in turbulent
astrophysical fluid flows, opening a wide field of applications involving feedback processes of young
massive stars on their parental clouds. However a self-consistent treatment for the thermal 
evolution was not included in 
this model and the energy evolution was approximated by a simple linear function of the 
ionisation ratio $x$,
i.e., $E=x E_{10000} + (1-x) E_{cloud}$, where $E$ is the internal energy of a representive particle with 
ionisation ratio $x$ ( $E_{10000}$  being the internal energy of a fully ionised particle
at a temperature of 10000 K and 
$E_{cloud}$ being the internal energy of a neutral particle at a temperature of 10 K). The absence of a
self-consistent treatment of the ionisation heating and recombination cooling processes in their model 
prevents a rigorous calculation of  
 the physical properties of the cloud, such as gas temperature, density, sound velocity and 
pressure \citep{Kessela,Kesselb}, which in fact can adequately 
be validated through  observational data.        
 
We therefore extend our previous code to include a wider range of UV radiation energy distributions,
 hydrogen ionisation, radiative transfer, ionisation heating and recombination cooling,
to provide a more comprehensive and self-consistent investigation on the effect of radiation of massive
stars on their parental  molecular clouds.  

\subsection{The main components in the previous SPH code}
A detailed description of the previous SPH code can be found in \citet{Nelson} and
\citet{White}. Here we present a  brief introduction of its main features for completeness. 
In the previous version of the code, the SPH numerical technique was employed to solve the following
continuity, momentum and energy equations for a compressible fluid
\begin{equation}
\frac{d \rho}{d t} + \rho \nabla \cdot \bf{v}  =  0 
\end{equation}
\begin{equation}
\frac{d \bf{ v}}{d t}  =  - \frac{1}{\rho} \nabla P - \nabla \Phi + \bf{S}_{visc}
\end{equation}
\begin{equation}
\frac{d \cal{U}}{d t} + \frac{P}{\rho} \nabla \cdot \bf{v}   =  \frac{\Gamma - \Lambda}{\rho}
\end{equation}      
and the chemical rate equations take the general form:
\begin{equation}
\frac{d X_i}{d t}=n K_i
\label{Xi}
\end{equation}
where $ \frac{d }{d t} = \frac{\partial}{dt} + \bf{v} \cdot \nabla$ denotes the convective derivative, 
$\rho$ is the density, $\bf{v}$ is the velocity, $P$ is the pressure, $\bf{S_{visc}}$ represents the viscous
forces, $\cal {U}$ is the internal energy per unit mass, and $\Phi$ is the gravitational potential. 
$\Gamma$ and $\Lambda$ represent nonadiabatic heating and cooling functions respectively.

The heating function $\Gamma$ in thermal modelling is mainly provided by the photoelectric emission of 
electrons from grains illuminated by the incident FUV ($6.5 < h\nu < 13.6$ eV) component 
at the surface of the cloud.
This is stronger on the surface facing toward the source star,  
 and  weaker on the shielded cloud surface since the latter is mainly from  backscattered light by 
the surrounding interstellar medium and is ~ 0.1-- 0.2 times that on the directly illuminated 
surface \citep{Desert,Hurwitz}, i.e., the FUV 
radiation flux incident the boundary of the cloud can be written as
\[  J_{FUV}(r =R, \theta) = \left\{ \begin{array}{cc}
                       2000 G_0 &  0^o \le \theta \le 90^o) \\
                       0.2*2000 G_0,  &  \, 90^o \le \theta \le 180^o)
                      \end{array} \right. \]                 
where $G_0$ is the standard interstellar UV field \citep{Habing} and $\theta$ is the azimuthal 
angle in  spherical coordinate. Heating of the gas is also affected by cosmic ray heating,
H$_2$ formation heating and gas-dust thermal exchange. 

The cooling function $\Lambda$ is affected by 
CO, CI, CII and OI line emission. The fractional abundance of  main chemical species $X_i$ in 
Equation (\ref{Xi}) are  for CO, CI, CII, 
HCO$^+$, OI, H$_e^+$, H$_3^+$, 
OH$_{x}$, CH$_x$, M$^+$ and electrons; $K_i$ is the associated chemical 
reaction rate and $n$ is the total  number density. 
Detailed formula for these physical and chemical processes can 
be found in papers of \citet{Nelson} and \citet{White}.

The dominant heating process included in the above is the photoelectric emission of electrons from
dust grains by the incident FUV (1--13.6 eV), as the code  was originally developed to 
observe the influence of the
interstellar radiation on the chemical and thermal evolution of an isolated  molecular clouds,
such as Bok globules. For the investigation of the influence of more 
intensive UV radiation from  massive stars            
on the evolution of its surrounding objects, hydrogen ionisation heating will play 
a dominant role since the intensive  Lyman continum photons available in the radiative flux,
 and the much higher abundance of hydrogen atoms than that of grain particles. Therefore it is 
 necessary to include the hydrogen ionisation process and the relevant heating and cooling
process in order to fulfill our goal.  
  
\subsection{New elements}
The new elements added into the above original code are solving for the ionising radiation transfer
equation, hydrogen ionisation heating and electron recombination cooling. We also introduce a
realistic description of the pressure of the molecular cloud by including the turbulent and magnetic
pressures. More detailed descriptions are contained in the following sub-sections.  
\subsubsection{Ionising radiation transfer}
Although Helium ionisation was included in the above chemical network, we could safely neglect it when 
dealing with the ionisation radiation transfer, ionisation heating and cooling for simplicity, because of
the much lower abundance of the Helium compared to hydrogen atoms \citep{Dyson}.  
The treatment of ionising radiation from nearby stars into the above
SPH code is based on solving the following ionisation rate and radiative transfer equations,
\begin{eqnarray}
\frac{d n_e}{d t}  & = & \mathcal{I} - \mathcal{R} \\
\frac{d J}{d z} &  = & - \sigma n (1 - x) J
\label{J0}
\end{eqnarray}
where 
$n_e = x n $ is the electron density by ionising hydrogen atoms, and $x$ is the ionisation fraction; 
${\mathcal{I}} ={\sigma} n (1 - x) J$ is the
ionisation rate, $J$ is the flux of Lyman continum photons at the site and $\sigma$ is the 
ionisation cross-section of hydrogen in the ground state.  
${\mathcal{R}} = n_e^2 \alpha_B = x^2 n^2 \alpha_B$ is the recombination rate, where $\alpha_B$ is the 
effective recombination coefficient under the assumption of the 'on the spot' approximation 
\citep{Dyson}. In its original definition, the recombination coefficient 
\[\alpha = \sum_i \alpha_i \] includes all of the individual recombination coefficients $\alpha_i$
to  the hydrogen atomic energy level $i$. In the 'on the spot' assumption, recombinations into the ground
level ($ i=1$) do not lead to any net effect on the change in ionisation rate, 
since the photons released from this recombination process  are able to re-ionise other hydrogen atoms 
on the spot. Therefore $\alpha_1$ can be neglected and the resulting net recombination coefficient can
be written as \[ \alpha_B = \sum \limits_{i=2}^{\infty} \alpha_i \]
The dependence of the recombination coefficient $\alpha_B$ on the temperature can be expressed by 
\cite{hummer},
\begin{equation}
\alpha_B = 1.627 \times 10^{-13} t_e^{-1/2}(1 - 1.657log_{10} t_e + 0.584 t_e^{1/3})
\end{equation}
where $t_e=10^{-4} T$ (T in K).  

The equations \ref{J0} is numerically integraded at each time step and with the density distribution 
$n_H(x,y,z)$ which resulted from the dynamical evolutions.where 
$J_0$ is the flux of Lyman continuum photons at the surface the 
molecular cloud and we take the value $J_0=1.5 \times 10^{11}$cm$^{-2}$s$^{-1}$, which is the same as that 
used by \citet{Williams}.
\subsubsection{Ionisation heating and recombination cooling}
The implementation of the thermal model for the ionisation process includes the energy input by 
photoionisation and energy loss by recombination. 
When an atom with ionisation energy $E_1 $ is ionised following the absorption of one photon of frequency
$\nu$ and energy $E = h \nu$, it releases one electron which carries the excess 
kinetic energy $ E_{\nu} - E_1$ which will be transfered to the gas through collisions with 
other gas particles. The resulting heating rate is described by the following equation 
\begin{equation}
\Gamma_{ionising}  =  n (1-x) \sigma J k T_* 
\end{equation}
where $k$ is the Boltzmann constant and $T_*$ is expressed as \citep{canto}
\begin{equation}
T_*=T_{eff}\frac{x_0^2 + 4 x_0 + 6}{x_0^2 + 2 x_0 + 2}
\end{equation}
where $T_{eff}$ is the stellar temperature, and $x_0 = (h\nu_0)/(k T_{eff})$ with $\nu_0$ being the
frequency of the Lyman limit.  

When a free electron in the plasma is captured by a proton, a photon is emitted and an amount of energy 
$E_1 + m_e v_e^2 / 2$ is removed from the internal energy of the gas. The cooling rate due to this
process is \citep{hummer}
\begin{equation}
\Lambda_{rec} = \beta_B \, n^2 x^2 k T 
\end{equation}
where $\beta_B = \alpha_B \times (1 + 0.158 t_e)$.
\subsubsection{Cooling by collisionally excited line radiaiton}
Accompanying with the ionisation of hydrogen atoms, oxygen is ionised to O$^+$ (OII) as well. 
Consequently, the collisional excitation of low-lying energy levels of OII
makes a significant cooling effect in spite of OII's low abundance, since OII has energy levels with 
excitation potentials of the order of $kT$.  We use the following simplified formula derived by \citet{raga} 
to calculate the cooling rate due to the collisional excitation of OII,  
\begin{equation}
\Lambda_{colli} = \Lambda_{colli \, (1)} + \Lambda_{colli \, (2)}
\end{equation}    
with
\begin{equation}
log_{10}[\frac{\Lambda_{colli \, (1)}}{n_e n_{OII}}] = 7.9 t_1 - 26.8
\end{equation}
\begin{equation}
log_{10}[\frac{\Lambda_{colli \, (2)}}{n_e n_{OII}}] = 1.9 \frac{t_2}{|t_2|^{0.5}} - 20.5
\end{equation}
with $t_1$ = $1 - 2000$ K$/T$ and $t_2 = 1 - 5 \times 10^4$ K$/T$ and $n_{OII} = x n_O$. 
The above formula is valid under the limit of low electron density $n_e < 10^4$ cm$^{-3}$, which 
is true for the BRCs systems. 

\subsubsection{The pressure of the cloud  and the boundary conditions}
\begin{figure}
\begin{center}
\resizebox{3cm}{6cm}{\includegraphics{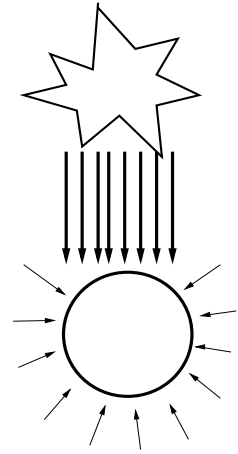}}
\caption{The initial geometric and the surrounding UV radiation configurations. The heavy arrow lines
represent the strong Lyman continum flux from the massive stars above the molecular 
cloud and the short arrow lines represent
 the isotropic FUV radiation ($ h\nu < 13.6$ \, eV) flux from the interstellar medium environment.}   
\label{geometric}
\end{center}
\end{figure}

At the start of a simulation, the molecular cloud is assumed to have a spherical symmetry 
which is the general assumption
adopted by  most  existing theoretical models as the initial morphology of the molecular cloud.
The geometrical evolution of an initially spherical molecular cloud under the 
effect of external environments may provide a platform for us to investigate the influence of 
the external radiation field on the morphology of the cloud. For 
investigation of the star formation at  the tips of Eagle Nebula fingers, it is the dynamical 
evolution of the head structure that we are interested in, therefore the above assumption on the 
initial shape of the simulated molecular cloud should be reasonable. The  mass  of the cloud  
is uniformly distributed through a sphere of radius $R$. In the simulation we choose the middle 
finger of the Eagle Nebula as an example since it has
a simpler structure than the other fingers \citep{White}. The initial geometric and  UV
radiation field configurations of the simulated cloud are shown in Figure \ref{geometric}.

For the head structure of the middle finger of the Eagle Nebula, with a mass $\sim$ 
 31 M$_{\sun}$ and a temperature of 20 K,  the Jeans length $R_{J} = 0.2$pc. However the radius of
the head structure is about 0.1 pc, which menas that the structure is unstable against
the self-gravity of the cloud according to Jeans criteria. A simple SPH simulation reveals that
it would collapse in $ < 10^5$years. 

However it is well known that molecular clouds are supported by turbulent and 
magnetic pressures in addition to the thermal pressure $P_{th}=n k T$. 
In the case of the Eagle Nebula, the observational results conclude that 
a large-scale ordered magnetic field is not likely to
to produce sufficient  internal pressure balancing the external pressure and  
there may be a disordered component of magnetic field which provides 
an isotropic pressure to balance the external pressure in the head structure 
\citep{White}. The relevant turbulent pressure $P_{tb}$ and the  
induced isotropic pressure $P_B$ due to magnetic field are related to the thermal 
pressure  by two dimensionless
parameters \citep{Gorti}, 
\begin{equation}
\alpha = \frac{P_{tb}}{P_{th}}; \, \, \, \, \, \, \, \, \, \, \, \beta = \frac{P_B}{P_{th}}
\end{equation} 

The internal pressure of the molecular cloud then consists of three terms:
\begin{equation}
P = (1+ \alpha + \beta ) P_{th}
\label{pressure}
\end{equation}

Observations of molecular clumps in star-forming regions \citep{Jijina}
indicate that the values of $\alpha$ range from about 0 in cold, dark clouds to $\ge 2$ in regions of 
massive star
formation. There is also observational evidence to suggest that turbulent support in clumps 
decreases on the smaller scales of star-forming clumps ($r_{c0} \le 0.2$pc), 
where $\alpha \le 2$ \citep{Goodman}. Therefore we adopt the value of $\alpha =2$
in our simulations considering the actual diameter of the condensed head at the tip of the middle 
finger is  0.2 pc. 

Measurements of magnetic fields and hence $\beta$ are difficult to make, but present 
observational data suggest a wide range of values for $\beta$, ranging from 0 to a few 
\citep{Crutcher}. We take $\beta$ as a parameter in our simulations and determine
a proper value of $\beta$ for the Eagle Nebula so that the total internal pressure can
balance the self-gravitation when the UV radiaition has not been switched on in order to
observe the effect of UV radiation triggered star formation in a molecular cloud. 
A value of $\beta=6$ is then obtained in this way.  

We start with a warm spherical cloud of mass 68 M$_{\sun}$ and temperature of 60 K. The initial radius
of the cloud is 0.4 pc and is assumed to have an uniform mass distribution and zero velocity distribution.
The boundary condition adopted is of constant external pressure  
which corresponds to an external medium composed primarily of atomic hydrogen with $n$(HI)=10 cm$^{-3}$ 
and $T=100$ K.

With these implementations to the original SPH code, we have developed a comprehensive three
dimensional SPH code of the radiation driven implosion effects of young stars on their
natal molecular clouds, which allows a detailed investigation on both dynamical and thermal evolution of 
a molecular cloud under the effects of nearby massive stars 
in a more accurate and realistic way 
than other existing models, because our model self-consistently includes most of the necessary
physical processes which are universal in astrophysical environment.

In the following subsections, we present and discuss the results from the simulations based on the 
above described  model with 20,000 simulated particles applied.

\section{Results and discussion}
\subsection{The dynamical evolution of the simulated molecular cloud}
\begin{figure*}
\resizebox{16 cm}{9 cm}{\includegraphics[0.4cm, 0.2cm][18 cm, 11cm]{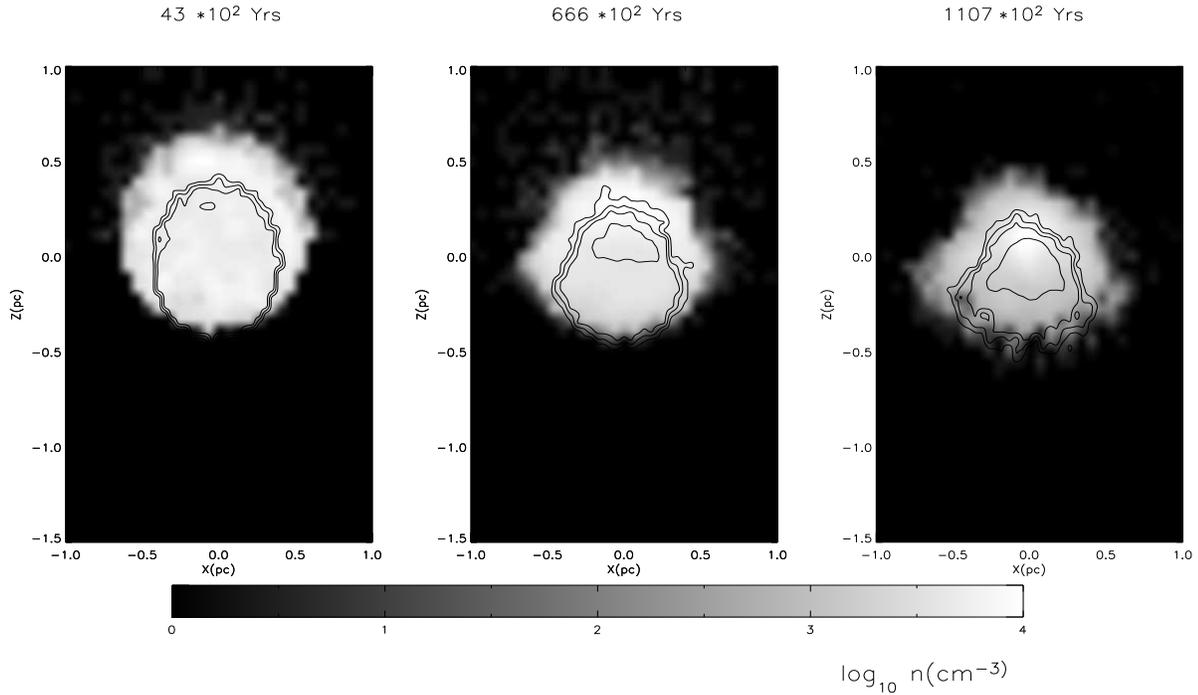}}
\caption{The evolution of the Hydrogen number density $n$ (cm$^{-3}$) of the simulated molecular cloud from 
early stages $t=4300$ year to  $t=0.11$ Myr, when a flat core has formed in the front part
of the cloud. The axis $z$ is along the strong radiation flux from the massive stars, and the $x$ axis 
is in the plane perpendicular to $z$. The initial center of the cloud is at $(x,z)$=(0.0)}.
\label{Eagle-1}
\end{figure*}

\begin{figure*}
\resizebox{16 cm}{9 cm}{\includegraphics[0.4cm, 0.2cm][18 cm, 11cm]{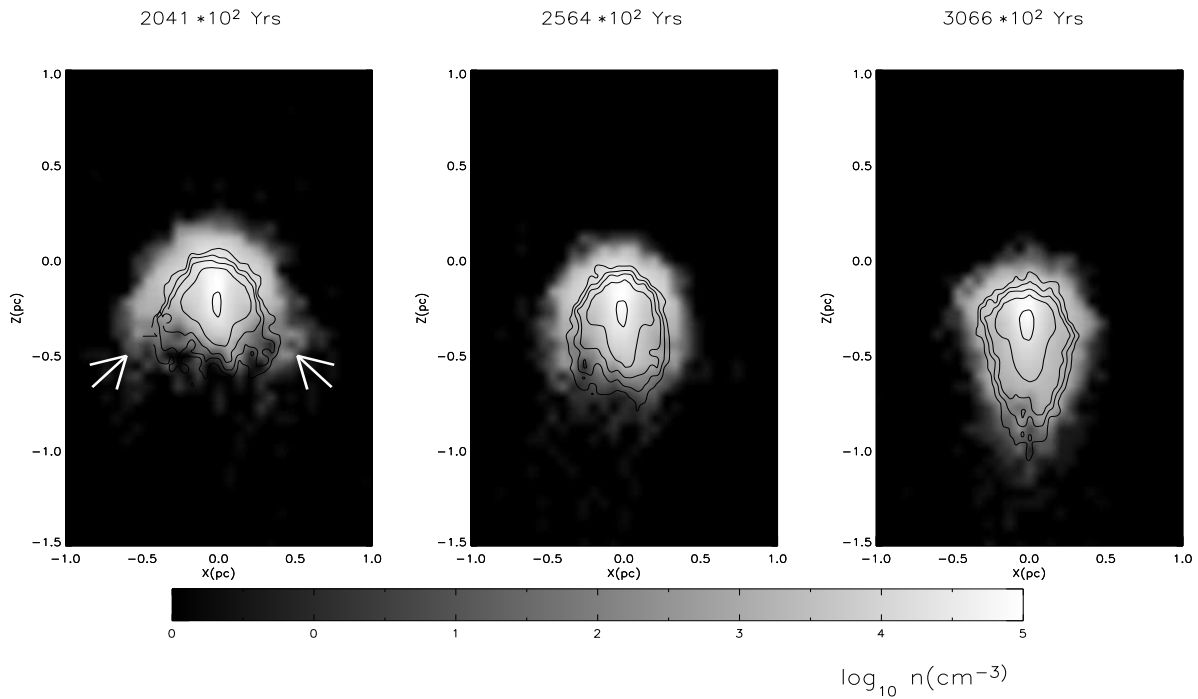}}
\caption{The evolution of the number density $n$ of the simulated molecular cloud over the period of 
$t=0.2 - 0.3$ Myr, when the core gradually becomes elongated. The white arrows point
to the place where the 'ear-like' structures are.} 
\label{Eagle-2}
\end{figure*}

\begin{figure*}
\resizebox{16 cm}{9 cm}{\includegraphics[0.4cm, 0.2cm][18 cm, 11cm]{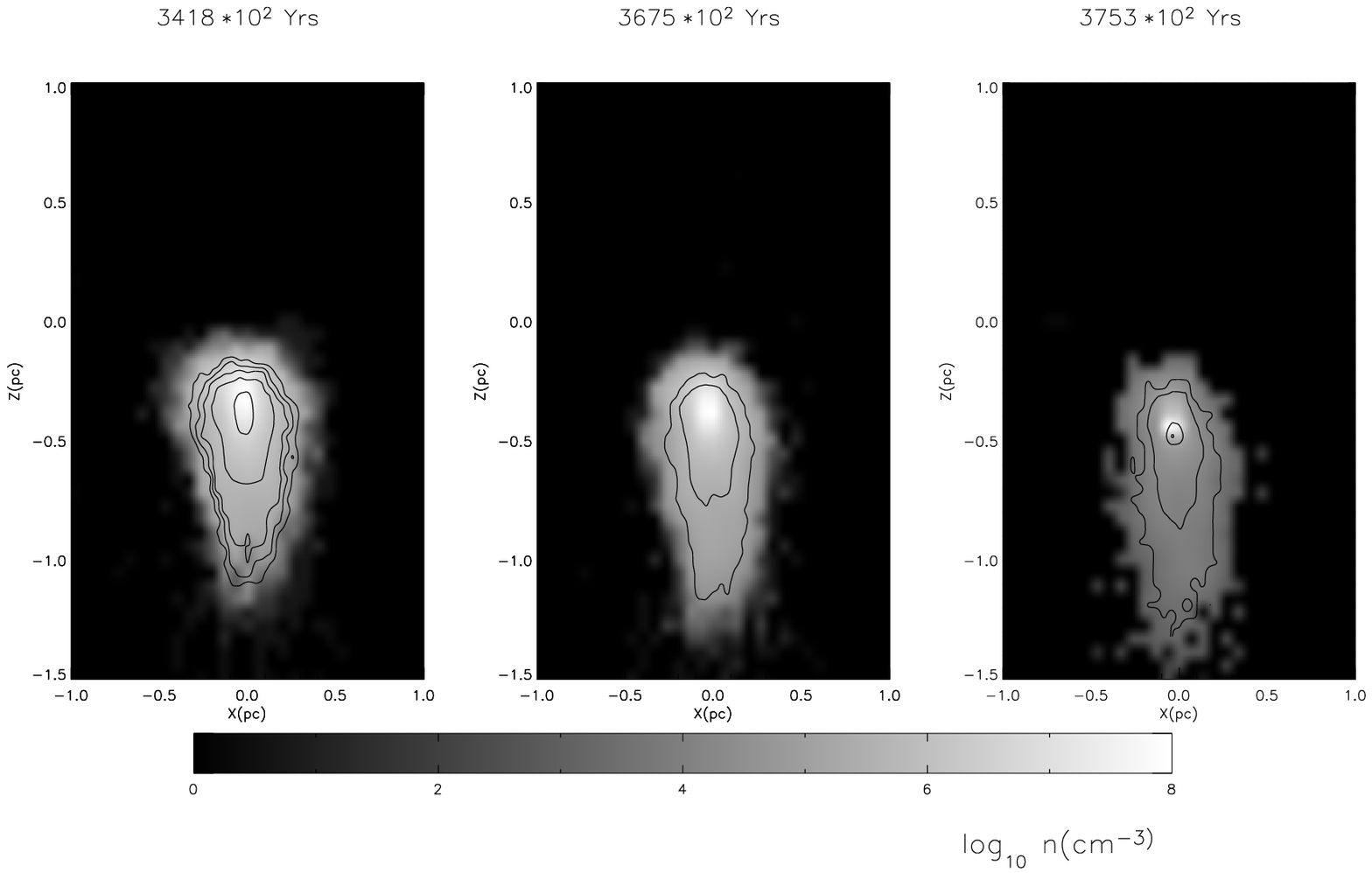}}
\caption{The final evolutionary stages of the condensed core collapsing into triggered star formation 
after 0.375 Myr.}
\label{Eagle-3}
\end{figure*}

Figures \ref{Eagle-1},\ref{Eagle-2} and \ref{Eagle-3} together show the number density evolution of
the spherical molecular cloud with the initial conditions stated in Section 3.2.3.
The overall evolutionary process can be categorised by three different stages according to the 
configuration of the simulated molecular core. In order to distingush different morphologies of the
core, we define a flat/elongated core as an object whose horizontal dimension is bigger/smaller
than the vertical dimension.


\begin{figure}
\begin{center}
\resizebox{8cm}{8cm}{\includegraphics{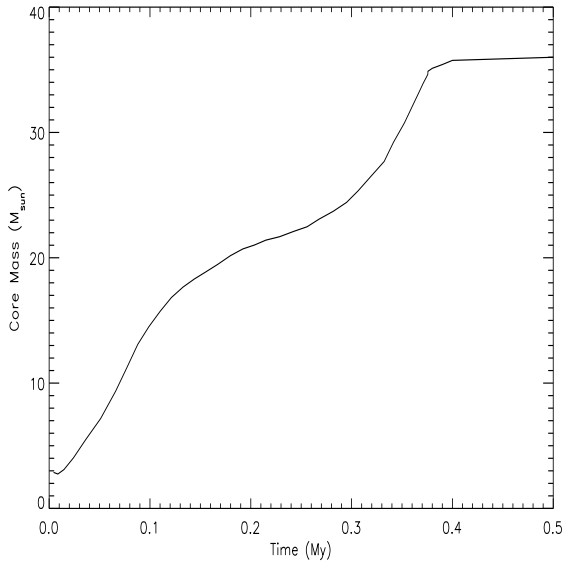}}
\caption{The Evolution of mass accumulated into the core of dimension $0.2\times0.2\times0.35$ pc$^3$ centered
in the densest point in the simulated molecular cloud.}
\label{core_mass}
\end{center}
\end{figure}

{\bf Flat core formation stage:} The left panel in Figure \ref{Eagle-1} shows the number 
density distribution of the simulated cloud at an early time $t=4300$ years. 
When the intensive Lyman continum flux from nearby massive stars falls onto the surface 
of the upper hemisphere 
of the cloud (front surface) as shown in Figure \ref{geometric}, 
the gas within a thin layer at the surface of the upper 
hemisphere is ionised and forms an ionisation front at the front surface. The ionised gas
is heated during the ionisation and the temperature increases.    
The increased pressure due to the temperature increase at this top layer 
drives an isothermal shock into the cloud, which compresses 
the neutral gas ahead of it (i.e., below the front surface). 
The contour lines at the front surface layer 
of the cloud in the left panel in Figure \ref{Eagle-1} show that a density gradient has 
built up there. The increased density 
in the neutral gas  leads to a rapid mass accumulation 
to form a condensed core. In  Figure \ref{core_mass},  
 the accumulated mass in a volume of  $0.2\times0.2\times0.35$ pc$^3$ ( similar
dimensions to the condensed core at the head of the middle finger of Eagle Nebula) centered 
at the densest point in the cloud, is shown as a function of time over the whole simulation period.        
The mass accumulation rate in the specified volume is about 93 M$_{\sun}$ My$^{-1}$ in the first few 
thousand years. On the other hand, the ionised and heated gas at the cloud surface flows radially 
away from the surface of the cloud 
and forms an  evaporating layer surrounding the  surface of the cloud,  
shown as the 'atmosphere' outside 
of these countour lines in the left panel of Figure \ref{Eagle-1}. 

The above stated shock compression and gas evaporation can be observed from the surface of the
rear hemisphere as well, but the effect is much weaker than that at the front hemisphere, because
 the heating effect of  photoelectric emission of electrons from dust grains is much weaker than
that of hydrogen ionisation by Lyman continum radiation.       

At time $t=0.666 \times 10^5$ years, the compressed layer (shock front) leading the ionisation front moves 
 toward the rear hemisphere,
while a flat core starts to form in the front hemisphere of the cloud due to the shock, as shown  
in the middle panel of Figure \ref{Eagle-1}. During the next  5$\times 10^4$ years, the core is continously
compressed and becomes flattened due to shock compression when  $t=1.107 \times 10^5$ years as shown in  
in the right panel of Figure \ref{Eagle-1}. 

With further compression of the core, as shown in the left panel of Figure \ref{Eagle-2}, 
the small 'ear-like' structures (indicated by the white arrows) similar to those 
in the  \citet{Lefloch} simulations  appear from the two sides of the cloud structure where the 
compressed layer meets the 
gas particles at the rear edge of the cloud at $ \theta \sim  90^o$. A small 'nose' structure also 
grows from the rear edge of the cloud along the central line $x=0$,  which is  
due to the maximum ionisation effect along the surface 
normal direction. Although the appearance of the 'nose' structure from the central line, 
the morphology of the cloud  and the core are still in a flat shape which is outlined by the contour lines in the left
panel of Figure\ref{Eagle-2}. In the next few $10^4$ years, the 'ears' gradually  merge  with 
 the 'nose' structure in the center and when  $t=2.564 \times 10^5$ years, 
the 'ears' finally disappear and  a flat core with a 'nose' structure along the center line, 
is shown in the middle panel in Figure \ref{Eagle-2}. Now the whole cloud is in a quasi-spherical morphology.
 In the section \ref{mechanism}, we will
present  a detailed discussion about the mechanism for the formation of the 'ear' and 'nose' structures. 

{\bf Elongated core formation stage:} 
In the next stage of the evolution of the simulated cloud, the whole  molecular 
cloud structure  gradually becomes 
elongated, as shown in the right panel of  Figure \ref{Eagle-2} when $t=3.066 \times 10^5$ years. 
When the time  $t=3.418 \times 10^5$ years, as shown in the left panel of
Figure \ref{Eagle-3}, a  finger structure appears. A  dense and elongated core
having a radius of $ ~ 0.1$ pc in the plane perpendicular to the ionising flux direction 
and length of 0.35 pc along the ionising flux direction forms as shown in the left panel of 
Figure \ref{Eagle-3}.  The mass included in the above core 
is about 31 M$_{\sun}$, and the average hydrogen number 
density is about $10^5$ cm$^{-3}$. The configuration of the simulated molecular cloud at this stage
 is very similar to the head structure of the middle finger  
observed by \citet{White}. We will further discuss  other physical features of the cloud
structure at this  moment of time in the following sections.
  
{\bf Triggered star formation stage:}
When $t=3.675 \times 10^5$ years, a highly condensed spherical core  starts to form at the head of the elongated 
structure due to the self-gravitation of the core, as shown in the middle panel of Figure \ref{Eagle-3}. With
further condensation of the gas into the center of the spherical core, the hydrogen density $n$ in the center 
of the core increases dramatically, reaching  $10^8$ cm$^{-3}$ when time $t=3.753 \times 10^5$ years.
The mass accumulated in the condensed elongated core ($0.2\times0.2\times0.35$ pc$^3$) 
reaches 34 M$_{\sun}$ when $t=3.753 \times 10^5$ year through  
compression by the isothermal shock induced by the ionising radiation, which causes the mass
to condense and then collapse into the center by gravitation, i.e., triggers the next generation of 
star formation in the center. The material in the condensed core quickly 
collapses into a smaller sphere of radius of 0.06 pc and a mass of about 25 M$_{\sun}$. 

The final mass of the whole finger structure is $~$ 60 M$_{\sun}$.  
In the whole evolutionary process, the molecular cloud has lost a mass of 8 M$_{\sun}$ throungh 
photoevaporation over 0.39 My, which results in an average mass loss rate of 
$\frac{8}{0.39} \sim 20.5$ M$_{\sun}$ My$^{-1}$. This value  is similar to  21 M$_{\sun}$ My$^{-1}$
which was obtained by  Lefloch's simulation \citep{Lefloch}. 
In the next section, we will discuss the mechanism for the   
appearances of  different morphologies of the cloud during the evolution of the cloud. 
\subsection{The dynamical evolution of the cloud structure}
By observing the changes in the morphology of the cloud structure, we are interested in the 
features of growth of the 'ears' and 'nose' structures  and also the time durations
for the cloud structures to stay in the 'flattened core' and 'enlongated core' stages.
 We will discuss them
by examining the evolution of the configuration of the velocity field shown in Figure \ref{velocity}, 
which reveals an
approximate evolutionary picture of the momentum of the cloud structure.      
\subsubsection{The growth of the 'ears' and 'nose' structures}
\label{mechanism}
\begin{figure*}
\begin{center}
\resizebox{14cm}{14cm}{\includegraphics{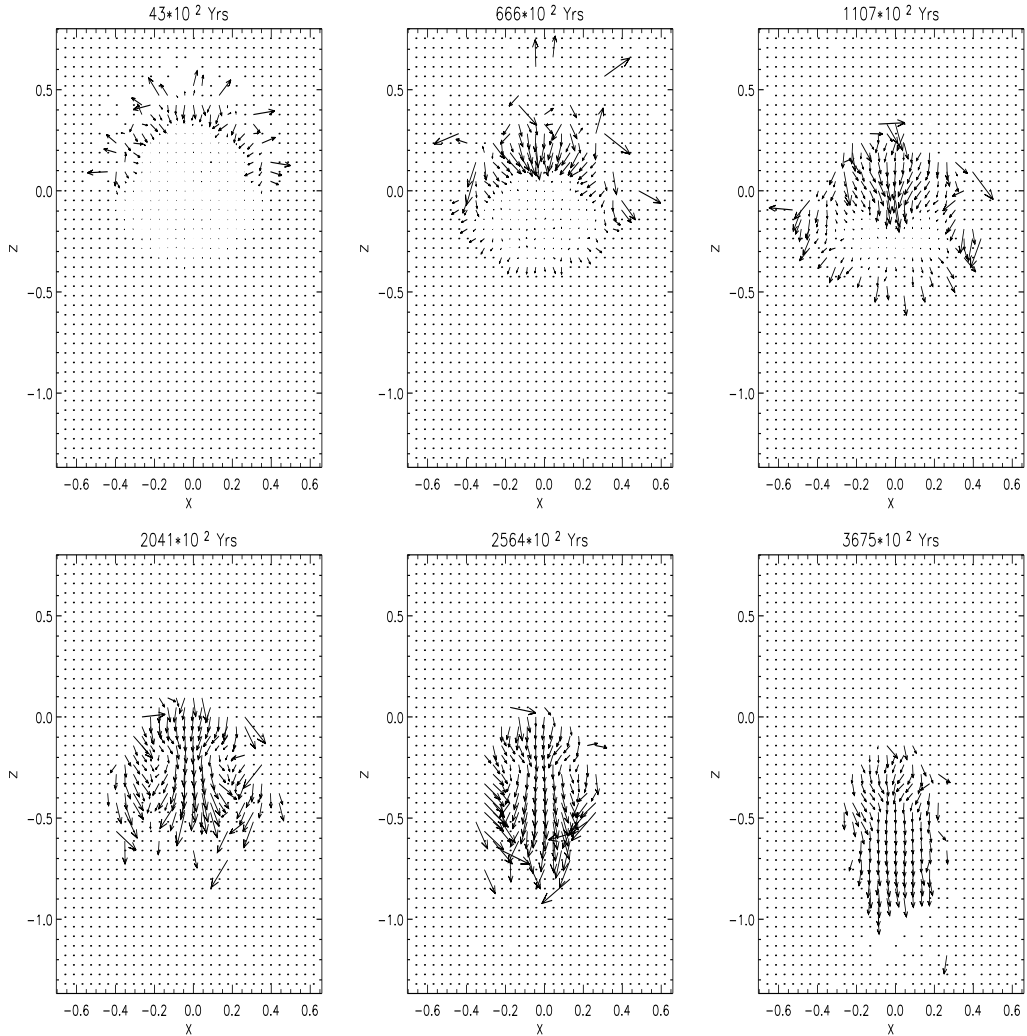}}
\caption{The velocity field evolution of the simulated cloud over a period time of 0.37 Myr. 
The length of the arrows indicates the magnitude of velocity and the longest arrow represents 
a velocitiy of +28 km s$^{-1}$.}
\label{velocity}
\end{center}
\end{figure*}

 At early evolutionary stages, while the ionised gas at the top layer of the front surface of the 
spherical cloud is heated and evaporated from the surface, an
 isothermal shock travels in the opposite direction towards the rear part of the cloud
 and compresses the neutral gas ahead, shown as
in the top left panel in Figure \ref{velocity}.    
The high pressure induced velocity distribution $V_z (\theta)$ at the front surface layer is approximately 
given by (Lefloch \& Lazareff, \cite{Lefloch}) 
\begin{equation}
V_z(\theta)= V_z (0)(cos(\theta))^{1/4}
\label{velo-e}
\end{equation}
where $\theta$ is the angle at a surface point $r=R$ from the z direction, and $V_z(0)$ is the velocity 
at the the point $r=R,\theta=0 $, which is 7.5 pc My$^{-1}$ (7.4 km s$^{-1}$) at the time $t=4300$ years. 
 The gas close to the symmetry axis ($x=0$) 
has a smaller $\theta$ at the front surface layer will move foward 
faster than gas which lies further away  (larger $\theta$ ) from the axis $x=0$. 
This velocity gradient over $\theta$ causes the deviation 
of the morphology of the front surface of the cloud from the hemisphere, i.e, the front 
surface of the hemisphere becomes squashed,  
which can been seen from the middle panel in Figure \ref{Eagle-1}.

The gas particles close to the outermost parts of the cloud ($\theta \sim  90^o$), as shown in the top-middle 
and top-right panels in 
Figure \ref{velocity} obtain an extra velocity along the z-direction due to the non-radial 
isothermal shock compression,
so that they  intersect the rear edge of the 
cloud and form  the 'ear-like' structures on the two sides of the cloud, which is shown in 
the left panel of Figure \ref{Eagle-2} 

Also the gas near the symmetry axis ($x= 0$) progresses forward to the rear hemisphere
much faster than that further from the symmetry axis due to its higher velocity 
obtained from the radiation induced compression, so that a 'nose' gradually grows out from the 
shock front along the $x=0$ axis over the next 0.44$\times 10^5$ years, 
dragging the most of the mass of the cloud with it,
as shown clearly in the right-top panel of 
Figure \ref{velocity}.   

In the following evolutionary stages, after the gas in the ears passes the front hemisphere 
they will merge into the moving nose structure along $x=0$ axis because the effect from the 
ionisation flux is 
greatly decreased inside the core region and the gravitational force of the nose 
structure becomes more important 
so that the gas in the 'ears' converges into the 'nose' structure. This effect can be seen from the 
bottom-left and bottom-middle panels of Figure \ref{velocity} by the 
directions of the velocities of the gas in the 'ears' pointing to the nose. 
The 'ears' merging with the moving 'nose'
gradually creates the tail structure of the cloud complex. 

The average velocity gradient in 
the dense core in the left panel of the Figure \ref{Eagle-3} (which corresponds to the current
state of the observed head structure of the Eagle Nebula) is  
$\frac{\partial v}{\partial r} \sim 1.88.$ km s$^{-1}$ pc$^{-1}$, which is consistent with 
 the observed value of 1.7 km s$^{-1}$ pc$^{-1}$ \citep{White}. 
The compressed gas left in the head of 
the cloud complex now  has reached a density which exceeds its Jeans density so that it  
collapses into a dense core  to form a new star there,  while the whole cloud complex 
keeps moving forward along the $-z$ direction due to the 'Rocket effect' \citep{Oort}
arising  from the evaporating gas leaving the front surface
of the cloud, as shown in the bottom-right panel of Figure \ref{velocity}. This 'Rocket effect'
makes the mass center of the cloud complex move a distance of 0.4 pc over a period of 0.37 My.  

\subsubsection{The duration of the flat and elongated morphologies}
Some of the above morphological features can be found in the \citet{Lefloch} model  as well, 
e.g., the 'ear' structures, the elongated cloud complex (or cometary ) morphology formed at the final 
stage of the evolution of the cloud. However we can not ignore some profound differences 
in the simulated morphologies
presented by these two models. Firstly in our simulations, the 'ear' structures does not last as long 
as in  Lefloch \& Lazareff's model, which we think may be the consequences of a) inclusion of gravitation in 
our model and b) the 3 dimensionlity of our model, so the gas in the 'ears' in our model converges  
 into the symmetry axis $x=0$ much earlier than that in Lefloch \& Lazareff's model (which is clearly seen 
by the directions of the velocities of the gas at two sides), which results in the fact that 
 the cloud complex stays longer at a flat-core morphology in our simulations . 
As can be seen from Figure \ref{Eagle-1}--\ref{Eagle-3}, the ratio of the time for the cloud 
complex staying in flat-core morphology to that in the elongated-core morphology 
is $\sim \frac{2.4}{3.9} \sim $62\%, which 
means that the simulated bright rimmed cloud spends 63\% of its 
lifetime in a type A (with a flat core) 
morphology, and $\sim \frac{3.9 - 2.4}{3.9} \sim $37\% of that in type B and C (with an elongated core), 
before it devolops further to form
cometary structure or collapse to  form a new star. 
We reach a similar conclusion from the results of a series of simulations performed for clouds of 
different masses. According to the statistical average principle, the above
results means that we should expected to   
 find twice as many of flat-core BRCs as enlongated-core BRCs when we observe a big group of BRCs in
a similar astrophysical environment. In comparison, the relevant 2-D modelling 
\citep{Lefloch,Leflocha} 
gave a value of $\sim \frac{0.126}{0.37} \sim $34\% for the time duration for the type A morphology.    

\begin{figure*}          
\setlength{\unitlength}{1cm}
\begin{picture}(14,6.5)\thicklines
\put(4,-0.){\vector(1,0){9}}
\multiput(4,2.5)(4,0){3}{\circle*{0.25}}
\qbezier(3,2)(4,3.5)(5,2)
\qbezier(7.3,0.5)(8.,5.2)(8.7,0.5)
\qbezier(11.8,0.5)(11.5,2.9)(12,3.)
\qbezier(12.,3.)(12.5,2.9)(12.1,0.5)
\multiput(3.5,5)(4,0){3}{\vector(0,-1){2}}
\multiput(4,5)(4,0){3}{\vector(0,-1){1.5}}
\multiput(4.5,5)(4,0){3}{\vector(0,-1){2}}
\put(3.5, 5.5){\framebox(1,0.5){Type A}}
\put(7.5, 5.5){\framebox(1,0.5){Type B}}
\put(11.5, 5.5){\framebox(1,0.5){Type C}}
\end{picture}
\caption{The schematic of three BRC morphologies. The vertical arrows represent the incoming ionising 
photon flux and horizontal arrow expresses the direction of the possible time-evolution of the cloud 
complex under the influence of the ionising radiation flux. The three types A, B and C correspond 
to the three morphologies in three different stages over the evolutionary of the cloud complex.}
\label{abc}    
\end{figure*}
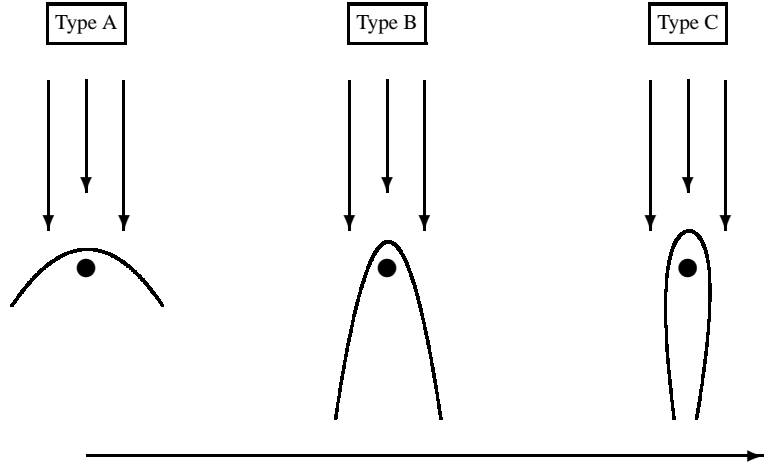

The above simulated morphological sequence of the cloud complex has been validated by many observations. 
\citet{Sugitani1} and \citet{Sugitani2} classified their observed 89 
BRCs (the SFO southern catalogue)
into types of A, B and C with an increasing curvature of clouds' rim,  shown as in the 
Figure \ref{abc}, which also roughly corresponds to the three stages in our simulated BRC evolutionary 
morphologies.It is found that the majority of BRCs (63\%) in the SFO southern catalogue 
are type A clouds, the remaining 37\% in  type B and C. 
Our latest radio continuum and molecular line observational study of  bright-rimmed clouds 
reveals a very similar BRC morphological distribution to that of Sugitani et al 
\citep{Urquhart,Urquhart2}.  A further investigation on the structure of BRCs,    
\citep{Ogura} supported by photoionisation induced shock models 
\citep{Vanhala}, also suggested
 that the above three morphological BRC types are possibly a time evolutionary sequence.    
Therefore the above interpretation of our simulated results on the morphological evolution of 
a BRC can be well supported by observations. The value of 63\% for the ratio of the number of 
flat-core BRCs to that of elongated-core BRCs is    
 consistent with the
observational value 66\%.   

In order to further assure the role of the gravity played in the triggered star formation process within
 the BRC's, we repeat the simulation without including gravity in the code, the result reveals  
similarities to those from \citet{Lefloch} and \citet{Williams} simulation, i.e., 
a) the core has a shorter time duration
in a flat core morphology than that with gravitation included; b) the core will not 
collapse and after 4-5 $10^5$ years
it stays in an equilibrium state with an average number denisty of 10$^5$ cm$^{-3}$ and
a central density up to 10$^6$cm$^{-3}$.  Therefore self-gravitation of the cloud 
indeed plays an important role in triggered star formation under the influence of 
 UV radiation field. We will present more simulation results and analysis 
 in our next paper for the discussion of the relationship between the morphology of the 
BRCs to cloud's initial structures.  
 
\subsection{The propagation of shock and the inonization front}
An important issue in the investigation of the  star formation 
within the Eagle Nebula fingers is to examine whether the propogation of the shock induced 
by the ionising radiation into the structure has triggered an irreversible collapse 
of the molecular cloud into a condensed center. From the simulated results, it is possible to
study the propagating process of the shock by observing the evolution of 
the number density distribution $n$ in the cloud structure.  

\begin{figure*}
\resizebox{17cm}{13cm}{\includegraphics{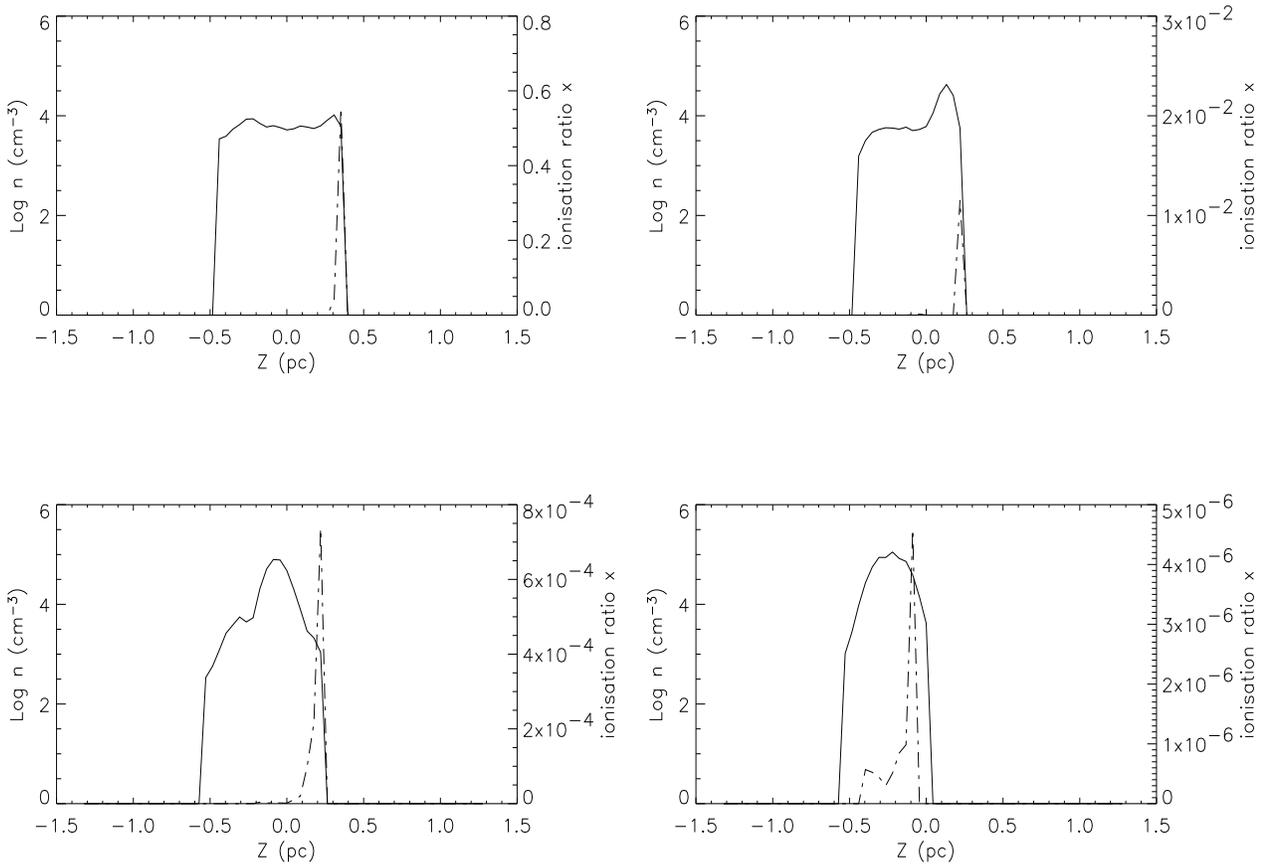}}
\caption{A shock (solid line) is leading the ionisation front (dot-dashed line) moving into 
the molecular cloud. From left to right, up to the bottom are the density distributions in 
the cloud along the symmetrical axis $x=0$ for the time t=0.0043, 0.067, 0.11, 0.2 Myr respectively.} 
\label{nx}
\end{figure*}

In Figure \ref{nx}, snapshots of the distribution of 
the number density $n$ and the ionisation ratio $x$ along the center line of the structure
at different times are shown. The left panel of the first row shows that at 
a very early stage ($t_1=0.0043$ Myr), 
the gas in the cloud is pushed towards the center not only from the front surface but also from
the rear surface due to the surrounding weaker UV radiation from the interstellar medium. 
The ionisation ratio $x$ at the front surface reaches a peak value of 0.58. Since the ionisation 
ratio $x$ is linearly related to the ionising flux $J_0$,  a
 sharp ionisation front is built up instantly. 
The hydrogen ionisation heating greatly increases the temperature
at the front surface which results in a high pressure so that a strong compressive shock 
is formed at the front surface of the structure, as shown in the right panel of Figure \ref{nx}. 
The bottom two panels show that the shock continues
leading the ionisation front into the cloud center at $t_3=0.11$ and $t_4=0.2$ Myr respectively.

Overall, the head of the finger becomes denser and denser under the effect of the ionisation 
induced shock compression. With the  increased density towards the center of the head, 
the ionising flux decreases very quickly ($J(z) \sim J_0 e^ {- \sigma n (1-x) z}$), 
which results in a rapid drop of ionisation ratio $x$ inside the head structure.   

If we define the region with highest density in the density profile as the shock front, 
the average speeds of the shock front  are $v_s = $ 0.33, 0.22 and 0.15 km s$^{-1}$
over the three periods  of $t_2 - t_1, t_3 - t_2$ and $t_4 - t_3$, while the corresponding
speeds for the ionisation front are $v_i =$ 0.18, 0.165 and 0.056 km s$^{-1}$ respectively. 
It is shown that
the propagations of both shock and ionisation fronts slow down because of the increased density
ahead. Since the ionisation front speed $v_i < v_s$, so the gap between the 
ionisation front and the shock front slowly increases.  After $t \sim 0.36$ Myr, the gap reaches 
a value of 0.125 pc, which is consistent with the fact that
the peak submm continum emission from the clumps is located at about 0.1 pc deeper into the fingers
than their photoionised front surface \cite{White}. 
In the following section, more details about this displacement will be discussed. 

\subsection{The evaporation of the ionised gases - HII region}
While the ionising radiation induced shock precedes the ionising front  into the cloud structure,
the ionised hydrogen atoms at the top layer of the upper-hemisphere are heated,  evaporated
and move away from the front surface of the cloud to form a HII region around the front 
surface of the cloud. Although the gas at the other sites of the surface other than the front sites 
are heated as well by the surrounding environmental FUV radiation through the photoelectric emision from
the surface of dust grains, the heating effect is much     
weaker (due to the low abundance of the dust grains) 
when compared with that of hydrogen ionisation so that the evaporating envelope is much thinner
than that surrounding the front surface, as shown in Figure \ref{Eagle-1}-\ref{Eagle-3}.

\begin{figure*}
\resizebox{17cm}{12cm}{\includegraphics{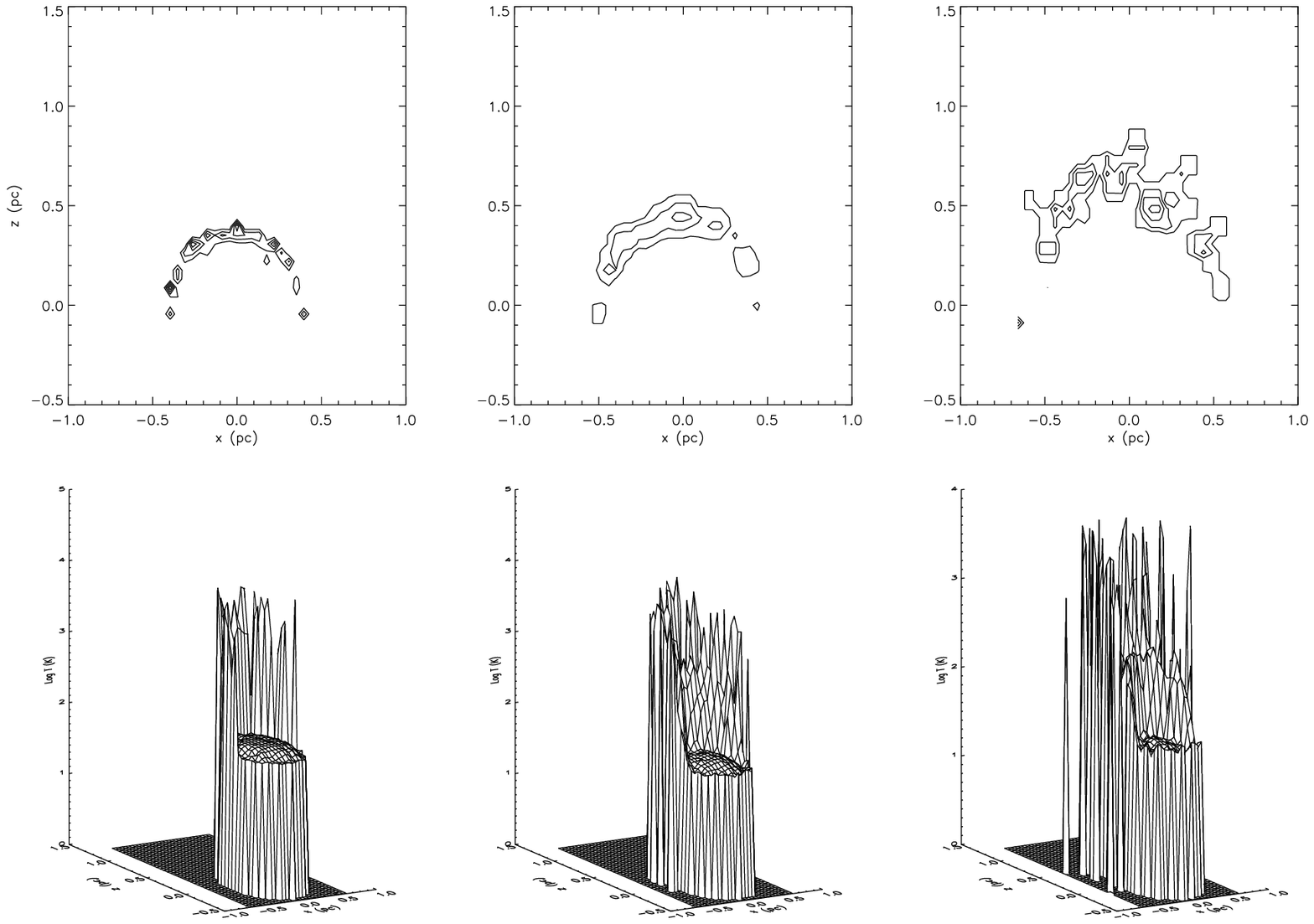}}
\caption{Top row: The ionised hot gas is leaving the upper semi-spherical surface. 
The contours are the values of
the ionisation rate $x$, the biggest value is 0.58. The left, middle and right panels are the 
ionisation rate distributions in the evaporating layer over the head of middle finger 
of the Eagle Nebula at times 0.0043, 0.009 and 0.015 Myr respectively.
Bottom row: The temperature (Log T (K)) distribution in the ionised hot gas over  the same period of times.}
\label{evaporation}
\end{figure*} 

 The panels in the upper row in Figure \ref{evaporation} show the movement of this 
evaporating layer for the first few  snapshots of $t$ = 0.0043, 0.009, and 0.015 Myr. The average 
evaporation speed is about 25 km s$^{-1}$. From the bottom  panels of the same figure, we can see 
that the  temperature of this evaporating layer is about several $ 6 \times 10^3 $ K, for which 
\citet{Hester} estimated a value of $10^4$ K, and \citet{White} gave a value of
$6700$ K. The characteristics of the modelled HII region surrounding the front surface of the
Eagle Nubula finger fits well with the observed values.    
\begin{figure}
\resizebox{7cm}{6cm}{\includegraphics{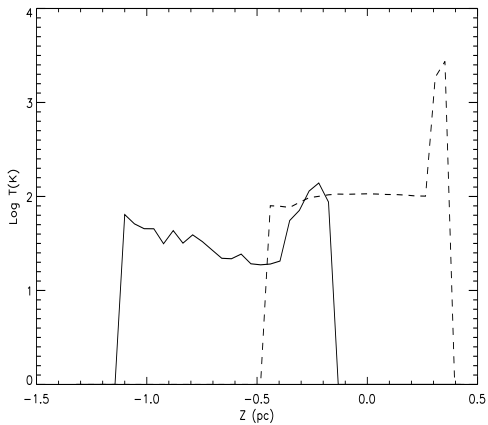}}
\caption{The temperature distributions at the initail and late stages of the evolution. Dashed line is for
$t=0.0044$ Myr, and the solid line is for $t=0.32$ Myr}
\label{Temp}
\end{figure}

\subsection{Evolution of the temperature profile}
The evolution of the temperature distribution of the head structure at  times $t=0.0044$
(the dashed line) and 0.34 Myr (the solid line) along the center line $X=0$, are shown 
in Figure \ref{Temp}.
It is seen that shortly after the ionising 
radiation is switched on, i.e., when  $t=0.0043$ Myr, the top layer at the front surface is heated 
as a consequence of hydrogen ionisation heating, so that the temperature increases to $\sim 6000$ K. 
Inside the structure the temperature is about
60 K, close to the inital temperature of the cloud, because the effect of the radiation 
has not yet penetrated. After the cloud has evolved for $t=0.32$ Myr, 
the temperature in the surface layer of the cloud
is about 200 K, since the ionisation flux has greatly decreased at the surface of the cloud due to 
the ionisation absorptions in the surrouding HII regions (which is moving away from the main structure)
for the recombinations of the electrons with the hydrogen
ions. The temperature at the rear surface increases to a similar value, due to the FUV radiation from
the surrounding interstellar medium. The combination of the value of the surface 
temperature ( $ \sim 200$ K) and the 
 hydrogen density ($n \sim 5 \times 10^4$) yields a  thermal pressure in  the surface 
layer of the $\frac{P_{th}}{k} = n T \sim 10^7$ cm$^{-3}$ K. 
The total pressure $\frac{P}{k} \sim 7 \times 10^7$
cm$^{-3}$ K according to Eq.({\ref{pressure}), which is in a  good agreement with the 
observed value by \citet{Pound}.    
The temperature in the core region of $x \times y\times z = 
0.2 \times 0.2 \times 0.35$ pc$^3$, which is centered 
at $(x,y,z)=(0, 0, - 0.43)$ pc, decreases to 13--18 K, below its original value of $T=60$ K, 
due to the enhanced radiation
extinction, and the increased cooling efficiency in this higher density region.           

In summary, the temperature distribution at the time $t=0.34$ Myr describes a low 
temperature (13 --18 K) core 
surrounded by an outer shell of warm material of temperature of 200-250 K, which matches well with the
the temperature characteristics of the oberved head structure at the middle fingertip in the 
Eagle Nebula \citep{White,Pound}.     
\subsection{The main chemical abundance evolutions}
The initial conditions for all of the chemicals abundances included are the same as those described in 
the previous paper, with $X$(C$^+$)=$X$(CO)=$X$(HCO$^+$)=0 and $X$(C)= $X(O) = 10^{-4}$ when the simulation
starts, where $X_i = \frac{n_i}{n(H_2)}$ \citep{White}. 
In the left panel of  Figure \ref{chemical}, the distributions of the fractional
aboundances of carbon bearing species are displayed for the time $t=0.32$ Myr.   
In the densest region 
between $z$=- 0.5 and - 0.3 pc, carbon  is mainly in the form of CO, which correlates with the 
density of H$_2$ because its production rate scale is $\propto n^2$(H$_2$), which makes the 
peak deep inside
the structure.  Outside the core $z > - 0.3$ pc, CI and C$^+$ are of comparable abundance
 at the front edge of the cloud strucure, due to an increased ionisation of CI into C$^+$ which 
results in a dramatic increasing aboundance of the latter. The peak of the ionisation ratio
$x$ at $z \sim - 2.7$ pc reflects the position of the bright rim. This is to say  that the ionisation peak
 is ahead of the CO peak by a distance
of ~ 1.25 pc, which provides a very consistent description of the relationship of the CO and H$_{\alpha}$ 
distributions inside the cloud head with the observed displacement of 0.11pc between CO line 
and H$_{\alpha}$ line profiles \citep{White}.   
 
The distribution of the HCO$^+$ abundance shows a non-monotonic variation over the region because
its production rate is dependent on the ionisation flux, the H bearing species and CO species. 
Near the front edge of the structure at $z=-0.3$ pc, these three factors 
makes a peak of the HCO$^+$ production rate. Deeper inside the structure, 
 the high abundance of CO is dominant in the production rate of HCO$^+$, therefore similar pattern
to that of CO, which is consistent with the observations of \citet{White}. 

White et al. also found  that the ratio of CI/CO abundance increases in 
the low column density regions. The right panel of the figure \ref{chemical} shows 
the corresponding ratio increases dramatically in the front region of the head structure, where the
density of the cloud is very low.  

\begin{center} 
\begin{figure*}
\resizebox{14cm}{7cm}{\includegraphics{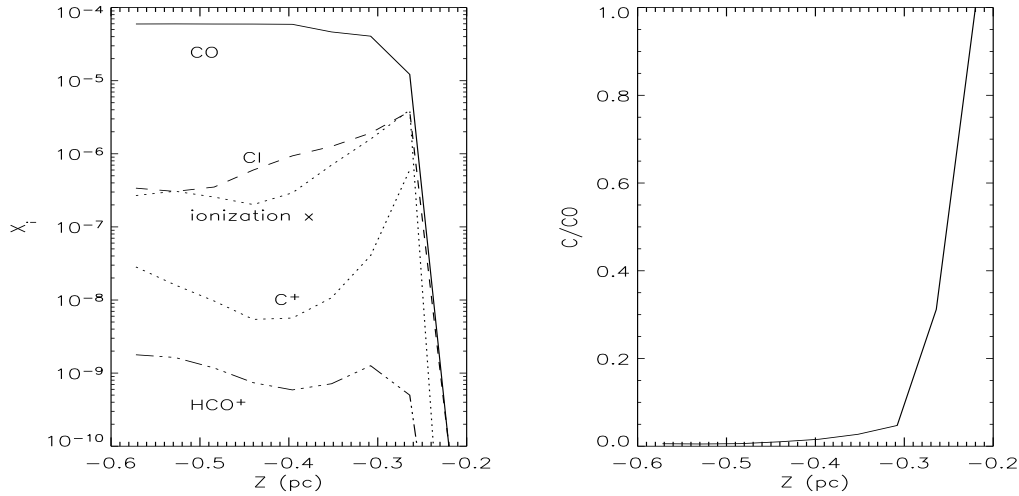}}
\caption{The main chemicals distributions at the time $t=0.34$ Myr}
\label{chemical}
\end{figure*}
\end{center}

\section{Comparison with the results from the previous model}
Firstly, we would like to point out that the estimated cloud thermal and chemical 
structures presented in the previous paper (\citep{White}) was based on a static
model under the assumption that the cloud was under hydrostatic equilibrium, i.e., 
the density distribution was fixed in time. Then the temperature and chemicals profiles 
were estimated according to the thermal and chemical models described in the previous
SPH code, which of course were not actually describe the dynamical evolution of the cloud
structure.

Secondly, a dynamical simulation based the previous SPH code 
in the absence of the magnetic and turbulent pressures support,  
indicates a quick collapse of an spherical molecular cloud of initial mass 68 M$_{\sun}$
and radius of 0.4 pc (or that of 31 M$_{\sun}$ and radius of 0.1pc, as we have used in our previous
paper \citep{White} in much less than 10$^5$ years 
which is not a reasonable result for the Eagle Nebula complex.       

Thirdly, as we have mentioned in the previous sections, the previous SPH code didn't  include
the Lyman continum in the radiation field, and the consequent 
hydrogen inoisation heating and cooling processes so it
greatly decreased the effect of radiative implosion on the bright
rimmed cloud, because the temperature near the bright rim of the cloud can only be increased to
190 K by the photoelectric effect of the FUV radiation. With the inclusion of Lyman continum
radiation, the ionising hearting can heat the front layer of the cloud up to \~ 7000 K in the first few 
hundred years, a reasonable temperature for HII region. 

Therefore the simulation based on the previous SPH code can only simulate the effect of photoelectric
heating of FUV radiation on the evolutionary of the molecular cloud, not the
whole evolution picture of how a ubiqutous molecular cloud under the effect nearby massive stars
 can evolve into a finger-like BRC with its life ending 
at either triggered star formation, formation of stationary cometary structure, or being evaporated totaly by
the nearby massive stars' radiative field. As we have shown in this paper, these effects can only 
be modelled via the comprehensively extended model.     

\section{Conclusion}
With the inclusion of the Lyman continum in the external radiation flux and hydrogen ionisation
heating effect into the energy equation in the previous code, 
and turbulent and magnetic pressures into the internal pressure of
molecular cloud, the upgraded 3D SPH model successfully describes the dynamical evolution of an initially
spherical uniform molecular cloud into cometary/triggered star formation under the 
influence of the strong UV radiation from nearby stars plus an isotropic surrounding interstellar medium
radiation, and self-gravitation of the molecular cloud. 
  
The application of the model to an initially spherical molecular cloud of mass 68 M$\sun$ 
and a temperature of 60 K results in an Eagle Nebula finger-like structure with a very 
dense core of a dimension 
of $0.2 \times 0.2 \times 0.35$ pc$^3$ and a mass of 31 M$\sun$ at its head after 0.34 Myr.   
The core has an average density of 10$^5$ cm$^{-3}$ and  a temperature 
of about 13-18 K with a velocity gradient of 1.91 km s$^{-1}$ at
time  $t=0.34$ Myr. The dense core is surrounded by a warmer layer of temperature 200-250 K. 
The simulated HII region is of temperature betweem 3-7000 K. 
The  charateristics of the simulated chemicals distributions are: a) CO abundance peak is 0.125 pc
deeper into the core from the peak of ionisation fractional ratio $x$ at the edge of the head structure; 
b)HCO$^+$ abundance follows a simillar pattern to CO in the core region; c)the abundance ratio 
of CI/CO increases with 
the decreaes of the cloumn density of the cloud. The above described characteristics of the 
simulated head structure coincide very well with the recent observational data \citep{White}. 
The simulation results also shows that the core will collapse into its center after another 7800 years.

It is concluded that the radiation induced compression wave has yet to
 pass the head structure and that the currently observed head structure at the tip of the middle finger 
is a part of an object which evolved from an initially larger and warmer pre-existing  
molecular clump. This cloud clump has been under the effect of a strong UV radiation from nearby stars 
 for about 0.34 Myr and
is currently at a transition stage towards triggered star formation. 

The simulation reveals some of the similar  morphological structures  in the evolutionary
process of the molecular cloud to those 
displayed by Lefloch \& Lazareff's 2-D model and obtains a very similar estimation on the
photoevaporation rate (20 M$\sun$ Myr $^{-1}$ vs 21 M$\sun$ Myr $^{-1}$ ) from the surface of 
the front side, which shows the reliability of our modelling. This also shows that the Lefloch \&
 Lazereff's 2-D model is appropriate to consider the  photoevaporation 
of a bright-rimmed cloud, but not the issue of triggered star formation in a the 
cloud. 

The estimation on the relative lifetimes of between type A and type B/C BRCs  
shows that a  BRC remains in type A morphology nearly twice the time spent in the 
type B/C morphology, which means that we should observe double the number of 
BRCs in Type A morphology. 
 This prediction agrees well with the observational results by  
\citet{Sugitani1} and \citet{Urquhart,Urquhart2}, thus provides a reasonable 
explanation for the observed unbalanced numbers of flat-core and elongated-core BRCs.

Comparing the above results with those based  the previous SPH model, we 
can conclude that the hydrogen ionisation heating by Lyman continum plays a much more important role 
in the dynamical evolution of a molecular cloud than that of photoelectric heating by the surrounding
interstellar medium radiation. Also the inclusion of the turbulent and magetic field pressures in the 
molecular cloud  in the upgraded model makes the model more realistic so that it can trace the origin of
an observed cloud, reveal the whole
dynamical evolution process of a molecular cloud to triggered star/cometary formation under the influence
of nearby stars, and predict the future evolution of the observed object. 
\section{Acknowledgements}
We would like to thank the anonymous referee for his/her helpful suggestions which makes
the presented work more reliable and also we learnt a lot from the communications with him/her.      

\end{document}